\documentstyle{mn}

\include{psfig}
\begin{document}

\title{Hot white dwarfs and the UV delay in dwarf novae}

\author[J.-M. Hameury, J.-P. Lasota and G. Dubus]{Jean-Marie Hameury$^1$,
	Jean-Pierre Lasota$^2$ and Guillaume Dubus$^2$ \cr
\\$^1$ UMR 7550 du CNRS, Observatoire de Strasbourg,
	11 rue de l'Universit\'e, F-67000 Strasbourg, France,
	hameury@astro.u-strasbg.fr
\\$^2$ UPR 176 du CNRS, D\'epartement d'Astrophysique Relativiste
	et de Cosmologie, Observatoire de Paris, Section de Meudon,
\\	F-92195 Meudon C\'edex, France, dubus@obspm.fr, lasota@obspm.fr}

\maketitle

\begin{abstract}
We calculate the effect of illumination of dwarf nova accretion discs
by radiation from a hot, central, white dwarf. We show that only for
very hot white dwarfs ($T_{\rm eff} \approx 40 000$~K) the inner region
of quiescent dwarf nova discs are partially depleted so that the delay
between the rise to outburst of the optical and UV fluxes would be increased
as suggested recently by King \shortcite{k97}. This depletion, however,
must create several small outbursts between main outbursts, contrary to
observations. Lower white dwarf temperatures may cause the outburts to
be of the `inside-out' type removing the UV delay. We conclude that
white dwarf irradiation of dwarf nova discs is not very efficient
for example because the UV radiation from the hot white dwarf does not
penetrate deep enough in the disc atmosphere.
The total ablation of the inner disc by e.g. evaporation (possibly related to
illumination)
appears to be a very promising possibility, accounting for both the EUV delay
and the general lightcurves properties.
\end{abstract}


\section{Introduction}

Dwarf novae (DN) are cataclysmic variables which exhibit large
amplitude outbursts that are commonly attributed to a thermal/viscous
instability occuring in the accretion disc around the white dwarf (primary)
component of the binary system (see e.g. Cannizzo \shortcite{can93} for a
review). In this picture, a stable disc has a bimodal behaviour: it can
be either in a hot, ionized, state, corresponding to large mass
transfer rates, or in a cool, neutral state, with a very low mass
transfer rate. Systems in which the mass transfer rate from the
secondary is in the intermediate range cannot be steady. The disc
becomes unstable when partial ionization of hydrogen sets in; this may
occur either in the innermost parts of the disc, in which case the
outburst is of the inside-out type, or in the outer disc (outside-in
outburst). In both cases, a heat front forms and rapidly propagates
across the disc, bringing it entirely (or almost entirely) to a hot
state, in which the local mass transfer rate is larger that the rate
at which matter is provided by the secondary. The disc thus empties
until a cooling wave starts from the outer edge of the disc, and brings
it entirely to the cool (and unsteady) state.

Whereas the disc instability picture accounts for the general observational
properties of DNs, it still has to face a number of contradictions, both
internal and external. Models suffer from a number of drawbacks and
inconsistencies \cite{lh98,hmdl98}. The description of viscosity which drives
accretion onto the white dwarf is purely phenomenological (and often {\sl ad
hoc}), so that it may be difficult to determine if failures to reproduce some
observational properties of dwarf nova outburst are due just to the
approximate treatment of the disc structure, or to some more fundamental
physical problems (see e.g. Gammie \& Menou 1998).

During some DN outbursts one observes a so called `UV-delay' - the rise to
outburst in the optical wavelength precedes the UV rise by as much as 0.5 to
0.75 day \cite{hps83,w95}; the delay between the optical and EUV can be as
long as 1 day \cite{mau96}. Such delays are observed during outburts which
start at the outer disc regions - the `outside-in' outbursts and are
usually assumed to be the propagation time of a heating front across the disc.
Whereas this interpretation can be questioned in the case of 1000 -- 2000 \AA
~UV, since the spectral modelling of the disc must be accurately done
\cite{can98}, the EUV rise clearly marks the arrival of the heating wave at
the white dwarf surface. The presence of
the UV-delay has been a serious difficulty for all disc models, since
according to them, the heating front propagates too rapidly across the disc -
the disc becomes hot too fast for an appreciable delay between UV and optical
to appear (see Fig. \ref{fig:delai}). Attempts to solve this problem by
modifying the local physical properties of the disc (e.g. increasing the
thermal scale -- Mineshige \shortcite{m88} -- or involving convection
--Duschl \shortcite{d89}) have not been successful. A much more promising
possibility was suggested by Meyer \shortcite{m90}: he proposed that the
innermost regions of the disc are ablated by a coronal siphon flow in
quiescence, so that UV emission, originating from the immediate vicinity of
the white dwarf, can occur only after the central disc has been
reconstructed, which is happening in a viscous time (see Hameury et al., 1997b
for the description of the same phenomenon in low mass X-ray binary systems).

Although the inability of DN models to reproduce observed UV-delays has been
attributed to the simplified way disc radiation is treated in time-dependent
calculations (see e.g. Cannizzo 1998) the EUV delay represents the real
heating-front propagation time because EUV radiation is not emitted by the
disc but by a boundary layer between the disc and the white dwarf.

The precise physical mechanism causing the truncation of the disc is
still a matter of debate. In addition to the evaporation of the inner
disc by siphon flow described by Meyer \& Meyer-Hofmeister
\shortcite{mm94}, Livio \& Pringle \shortcite{lp92} suggested that the
inner hole is due to the presence of a weak ($\sim 10^4$ G) magnetic
field. It is interesting that Patterson et al., (1998) found that the
white dwarf in WZ Sge may have a magnetic field $(1-5) \times 10^4$ G
and that the presence of a truncated disc in this system had been
proposed before by Lasota, Hameury \& Hur\'e (1995) and Hameury, Lasota
\& Hur\'e (1997a).

In a recent paper, King \shortcite{k97} argues that irradiation by the
white dwarf also leads to the truncation of the disc. According to King
\shortcite{k97} disc regions close to a hot enough white dwarf cannot
be brought to a cool state at the end of a DN outburst, so that matter
is transferred rapidly onto the white dwarf, leading to the formation
of a very low density, hot flow.

The effect of disc irradiation in the context of soft X-ray transients
was studied by Tuchman, Mineshige \& Wheeler \shortcite{tmw90}, and may
play an important role in low-mass X-ray binaries in general
\cite{vp96,dlhc98}. In this paper, we study the effect of the
illumination of the disc by the white dwarf. We show that, as proposed
by King \shortcite{k97}, the inermost disc regions are depleted, but
contrary to his hypothesis they still remain optically thick. The main
effect of this illumination is to render the transition region between
the hot and cold regions extremely unstable. As a consequence, many
small outbursts are triggered, outside-in outbursts are difficult to
obtain, and most often, the paradoxical result of illumination is a
suppression of the UV delay. In Dubus et al. \shortcite{dlhc98} we present a detailed
analysis of the case in which the illuminating flux originates from the
disc itself, and is therefore coupled to the mass accretion rate onto
the compact object; in the present study we find that adding
irradiation due to accretion does not modify our main conclusions.

\section{Modification of the $\Sigma - T_{\rm \lowercase{eff}}$ curves by
irradiation}

The main effect of disc irradiation from the outside is to change the surface
temperature. A fraction $1-\beta$ of the incident flux is absorbed in
optically thick regions, thermalized and reemitted as photospheric radiation.
The remaining is either scattered, or absorbed in optically thin regions,
possibly forming a warm corona \cite{ish96}. Because the white dwarf
temperature is not very different from that of the disc surface, King
\shortcite{k97} assumed that the albedo $\beta$ is not close to unity - in
this case effects of irradiation on the inner disc structure can indeed be
important.

In the case of a geometrically thin disc surrounding a white dwarf whose
radius is larger than the disc thickness, the illumination temperature
$T_{\rm ill}$ defined as
\begin{equation}
\sigma T_{\rm ill}^4 = F_{\rm ill}
\end{equation}
where $F_{\rm ill}$ is the radiative flux illuminating each side of the disc
is given by (Friedjung 1985; Smak 1989; Hubeny 1991):
\begin{equation}
T_{\rm ill}^4 = (1-\beta) T_*^4 {1 \over \pi} [\arcsin \rho -\rho
(1-\rho^2)^{1/2} ]
\label{eq:till}
\end{equation}
where $\rho = R_*/r$, $R_*$ and $T_*$ are the white dwarf radius and
temperature, and $r$ is the radial coordinate in the disc.

\begin{figure}
\psfig{figure=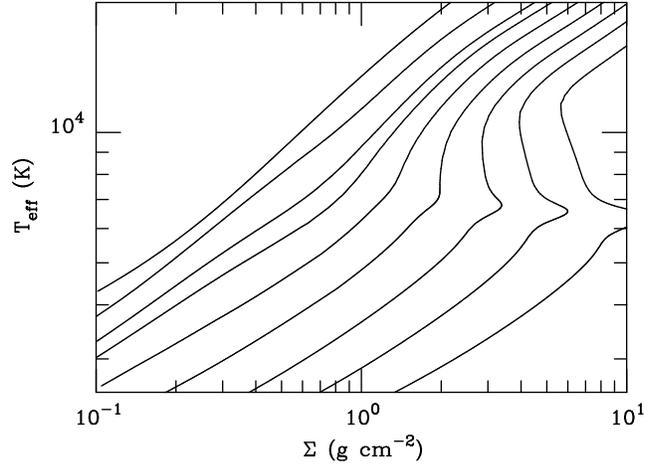,width=\columnwidth}
\caption{$\Sigma- T_{\rm eff}$ curves for $\alpha = \alpha(T,r)$ (see text),
a 0.6 M$_\odot$, $8.5 \times 10^{8}$ cm radius white dwarf, with
$(1-\beta)^{1/4} T_*$ = 25,000 K, for radii 0.85, 1.09, 1.29, 1.39,
1.54, 1.71, 1.90, 2.11, 2.42 $\times 10^9$ cm from left to right. Note that
$T_{\rm eff}$ accounts only for viscous dissipation and differs from the
actual surface temperature.}
\label{fig:sigteff}
\end{figure}

We have solved the disc vertical structure in the optically thick
approximation, as described in Hameury et al. \shortcite{hmdl98}, having
modified the disc outer boundary condition according to:
\begin{equation}
\sigma T_{\rm s}^4 = F_{\rm visc} + \sigma T_{\rm ill}^4
\label{eq:ts}
\end{equation}
where $T_{\rm s}$ is the disc photospheric temperature and $ F_{\rm visc}$
is the energy flux due to viscous dissipation. Examples of the
resulting $\Sigma - T_{\rm eff}$ curves are given in
Fig.~\ref{fig:sigteff}, where $\sigma T_{\rm eff}^4 \equiv F_{\rm
visc}$ (see e.g. Dubus et al., \shortcite{dlhc98}), which is
proportional to the mass transfer rate in steady state. We have assumed
a bimodal behaviour of the Shakura-Sunyaev parameter $\alpha$, taken
as:
\begin{eqnarray}
\log (\alpha)=\log(\alpha_{\rm cold})& +& \left. \left[ \log(\alpha_{\rm hot})-
\log( \alpha_{\rm cold} ) \right] \right/ \nonumber \\ & &
\left[1+ \left( \frac{2.5 \times 10^4 \; \rm K}{(T_{\rm c}^2 + 5 T_{\rm
ill}^2)^{1/2}} \right)^8 \right] ,
\label{eq:alpha}
\end{eqnarray}
where $T_{\rm c}$ is the central disc temperature, and $\alpha_{\rm cold}$
and $\alpha_{\rm hot}$ are constants, here taken to be 0.04 and 0.2
respectively. Note the presence of a $T_{\rm ill}$ term, that did not appear
in Hameury et al. \shortcite{hmdl98}, and which is needed to have $\alpha =
\alpha_{\rm hot}$ on the hot branch for moderate $T_{\rm c}$.
Fig.~\ref{fig:sigteff} shows the the `effective S-curves' used in
time-dependent calculations and not the constant-$\alpha$ S-curves that are
usually found in the literature (see also Hameury et al. \shortcite{hmdl98})

\begin{figure}
\psfig{figure=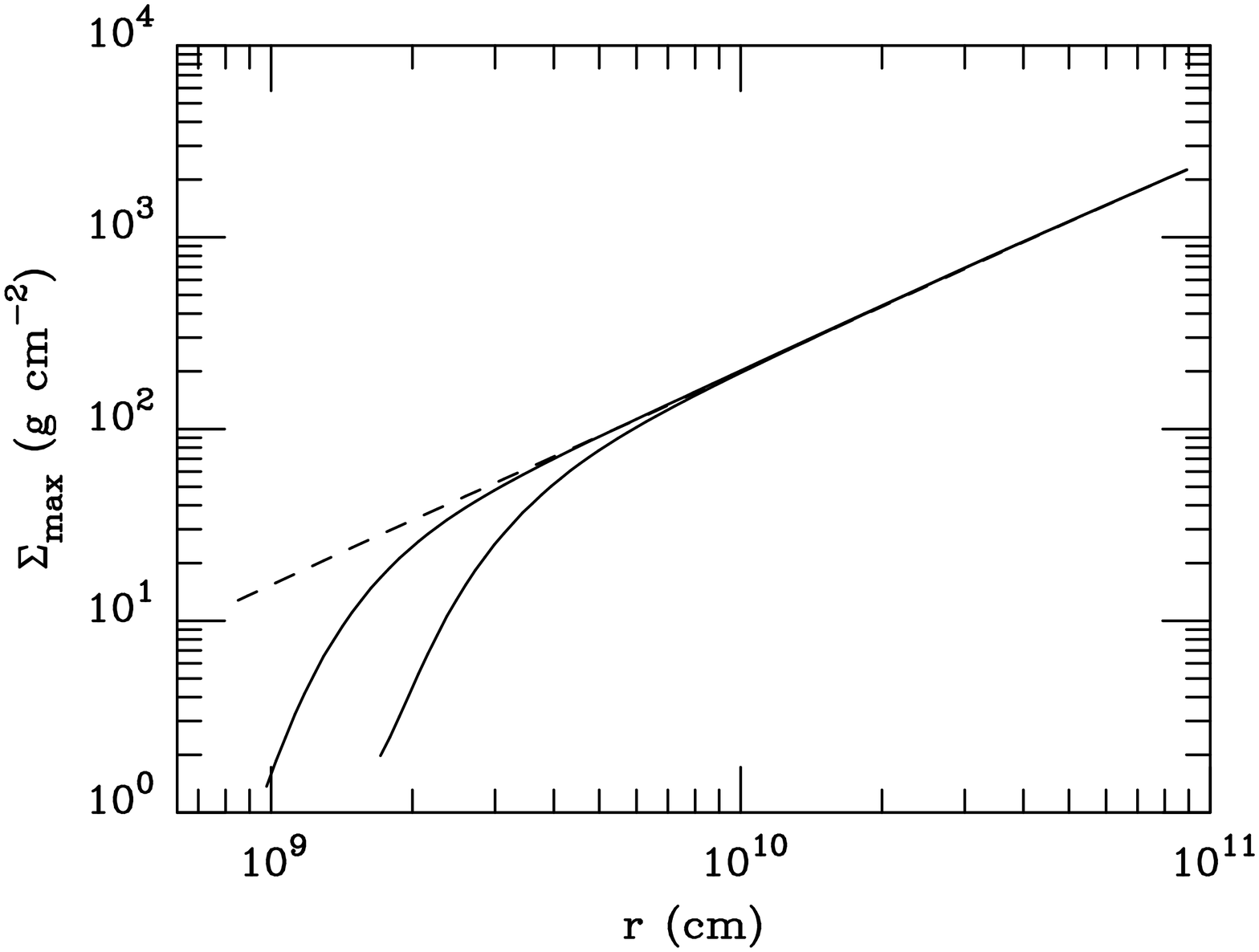,width=\columnwidth}
\caption{Maximum value of $\Sigma$ on the lower stable branch of the $\Sigma
- T_{\rm eff}$ curve. The primary is a 0.6 M$_\odot$, $8.5 \times 10^8$ cm
white dwarf, with effective temperature 25,000 K (lower solid curve), or
15,000 K (upper solid curve). The unilluminated case (dashed curve) is also
shown for comparison.}
\label{fig:smax}
\end{figure}

As can be seen, regions in the immediate vicinity of the white dwarf are
stabilized by irradiation from the white dwarf when the central disc
temperature is pushed above the hydrogen ionization range. The condition for
this to happen is close to that given by King \shortcite{k97}, i.e. $T_{\rm s}
> 6500$ K as in unilluminated discs, but with $T_{\rm s}$ now given by
Eq.~\ref{eq:ts}; this condition slightly overestimates the effect of
illumination, as the relation between $T_{\rm s}$ and the central disc
temperature is altered.
This is however not the sole
effect of irradiation. Figure \ref{fig:smax} shows the corresponding maximum
value of surface density on the cool branch, $\Sigma_{\rm max}$, as a
function of radius, $\alpha$ being given by (\ref{eq:alpha}). 
On Fig. 2 the curves describing maximum allowed $\Sigma$ for cold, stable, disc
equilibria end at radii where the disc become stable (there are no more
`S-curves' for smaller radii). At larger radii, however, irradiation has a
strong destabilising effect as seen in Fig. 2 where $\Sigma_{\rm max}$ is
reduced by more than an order of magnitude compared to the non-irradiated case.
In these regions, irradiation increases the midplane temperature, bringing it 
close to the hydrogen partial ionization regime.
Equivalently, the maximum mass transfer rate on the lower
branch is severely reduced. This, as we will see in the next
section, has an important influence on the properties of dwarf nova outbursts
when they are affected by irradiation from the white dwarf.

\section{Light curves and disc irradiation}

\begin{figure}
\psfig{figure=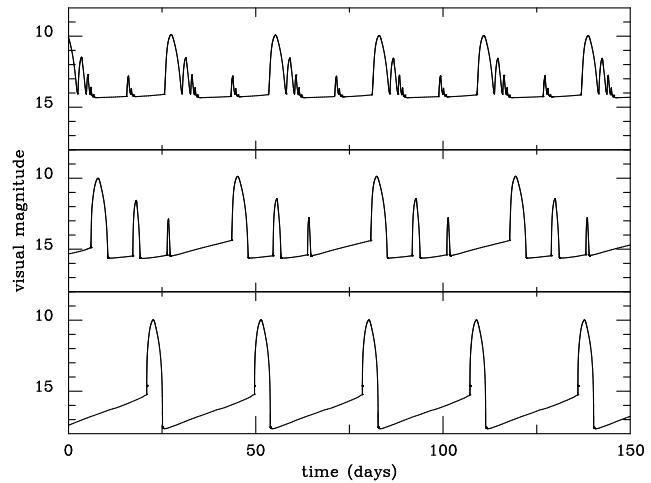,width=\columnwidth}
\caption{Visual light curves of the accretion disc with a 0.6 M$_\odot$, $8.5
\times 10^{8}$ cm radius white dwarf, $\dot{M} = 10^{16}$ g~s$^{-1}$ with
$(1-\beta)^{1/4} T_*$ = 25,000 K (upper panel), 15,000 K (intermediate
panel), and no illumination (lower panel). The contribution of the primary or
the secondary is not taken into account}
\label{fig:prof1}
\end{figure}

\begin{figure}
\psfig{figure=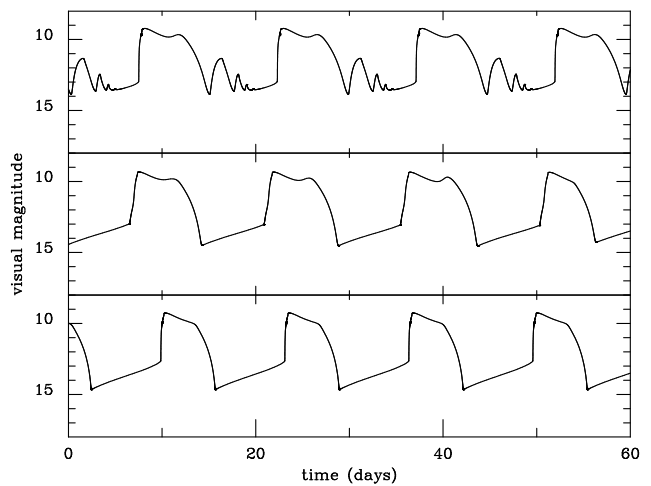,width=\columnwidth}
\caption{Same as Fig. \ref{fig:prof1}, but for $\dot{M} = 10^{17}$
g~s$^{-1}$. Note that the scale of the time axis are quite different}
\label{fig:prof10}
\end{figure}

We calculated a grid of vertical structures modified by irradiation. Such a grid
provides the cooling term as a function of the central temperature and surface
density and we we used it in the numerical code described in Hameury et al.
\shortcite{hmdl98} to determine the time evolution of the outbursts. Figures
\ref{fig:prof1} and \ref{fig:prof10} show the general light curve of a system
with a 0.6 M$_\odot$, $8.5 \times 10^8$ cm white dwarf surrounded by a disc
with an average outer radius $2 \times 10^{10}$ cm, fed at a rate of
$10^{16}$ and $10^{17}$ g~s$^{-1}$ respectively, for three different values
of $ (1-\beta)^{1/4} T_*$: 0, 15,000 and 25,000 K. In the unilluminated case,
inside-out outbursts are obtained at low mass transfer rates, whereas one
gets outside-in outbursts for $\dot{M} = 10^{17}$ g s$^{-1}$.


The first visible effect of illumination is the appearance of intermediate,
smaller outbursts that are always of the inside-out type; these appear
because of the destabilization of the inner edge of the disc by irradiation
as seen on Fig. \ref{fig:smax}. They are at least one magnitude fainter than
the major bursts, but are still detectable. During these small outbursts,
the heat front is not able to propagate across the entire disc, but dies out
at rather small radii (typically less that one half of the disc radius). The
characteristics of the main outbursts are not drastically changed, with one
exception. With $\dot{M} = 10^{17}$ g s$^{-1}$ and no illumination, the
outbursts are of the outside-in type, as noted above. With $T_*$ = 25,000 K,
the major outbursts are still of the outside-in type, while the smaller ones
are inside-out. There is a transition regime when $T_*$ = 15,000 K where only
large inside-out outbursts are seen. For this intermediate regime, the
surface density is everywhere close to $\Sigma_{\rm max}$ in the cold state
(as in the non-illuminated case) but illumination is strong enough to
destabilize regions close to the white dwarf. Heat fronts created in the
inner regions of the disc can thus propagate throughout the disc and not only
in a limited region. In other words, for these parameters the `small'
inside-out outbursts become major outbursts. These light curves are very
similar to those obtained with higher primary masses, and hence much shorter
inner disc radii, in which small inside-out outbursts are also seen
\cite{can93,hmdl98}.

A second effect is the reduction of the outbursts amplitude. They are
only slightly fainter at maximum, but the main contribution to the
reduced amplitude comes from the increased brightness during
quiescence, that results from the brightening of the inner parts of the
disc due to the reprocessing of light from the white dwarf.

\begin{figure}
\psfig{figure=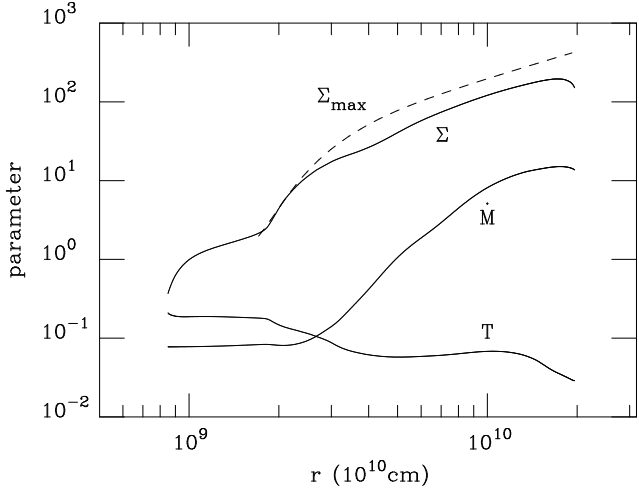,width=\columnwidth}
\caption{Radial profiles of temperature $T$ (in 10$^5$ K), surface density
$\Sigma$ and critical surface density $\Sigma_{\rm max}$ (in g cm$^{-2}$ and
local mass transfer rate (in units of 10$^{15}$ g~s$^{-1}$ just prior to one
of the major outbursts shown in Fig. \ref{fig:prof1} for $(1-\beta)^{1/4}
T_*$ = 25,000 K.}
\label{fig:rprof}
\end{figure}

\begin{figure}
\psfig{figure=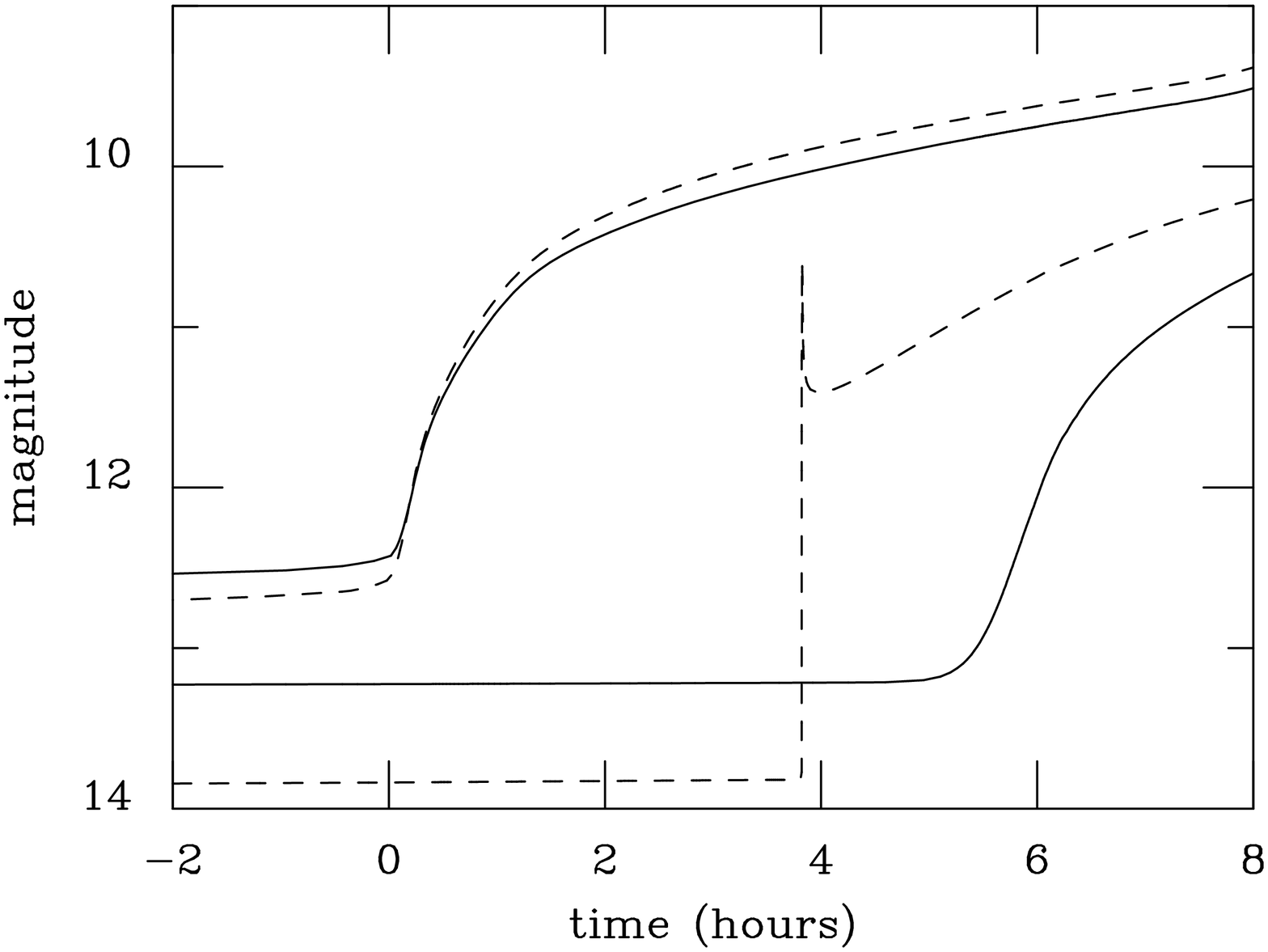,width=\columnwidth}
\caption{Time dependence of the V magnitude (left two curves) and the
"accretion rate magnitude" (right two curves, see text) at the onset of
an outside-in outburst under the
effect of illumination (solid curve, $(1-\beta)^{1/4} T_*^4 = 25,000$ K) and
without illumination (dashed curve). The primary mass is 0.6 M$_\odot$ and
the mass transfer rate is $10^{17}$ g s$^{-1}$.}
\label{fig:delai}
\end{figure}

As suggested by King \shortcite{k97}, illumination does reduce the surface
density in the inner parts of the disc. Fig. \ref{fig:rprof} shows the radial
disc structure in a system with a 0.6 M$_\odot$, $8.5 \times 10^8$ cm white
dwarf, $\dot{M} = 10^{16}$ g~s$^{-1}$, and $(1-\beta)^{1/4} T_*$ = 25,000 K
just prior to a major outburst. $\Sigma$ is much reduced in the region where
illumination is important, i.e. up to about 2 white dwarf radii, but not to
the point making the disc optically thin. For the value of $\dot{M}$
considered here, the outburst will be of the inside-out type, as can be seen
from the relative position of the $\Sigma$ and $\Sigma_{\rm max}$ curves. For
higher mass transfer rates, outside-in outbursts are possible. Fig.
\ref{fig:delai} shows the time behaviour of the onset of the bursts of both
the disc visual magnitude, and the ``accretion rate magnitude", defined as
$m_x = 27 -\log \dot{M}_{\rm wd}$, where $\dot{M}_{\rm wd}$ is the accretion
rate onto the white dwarf in g~s$^{-1}$, which represents the EUV flux. As
expected, the delay between the start of the rise of the disc brightness and
the rise of mass accretion rate onto the white dwarf increases when
illumination is important. The increase is however not very large; for the
parameters assumed here, the total delay amounts to about 5.3 hr, of which
about 1.5 hr is attributable to the depletion of the inner disc; longer
values are possible for higher white dwarf temperatures, or lower values of
$\alpha$.

It is also interesting to note that illumination makes the rise of the mass
accretion rate onto the white dwarf much less sudden than in the
unilluminated case. This is due to the fact that close to the white dwarf,
the heat front propagates now in a hot disc, with a constant value of
$\alpha$. This reduces the non-linearities which are responsible for the
presence of sharply defined heat fronts.

\section{The case of SS Cyg}

\begin{figure}
\psfig{figure=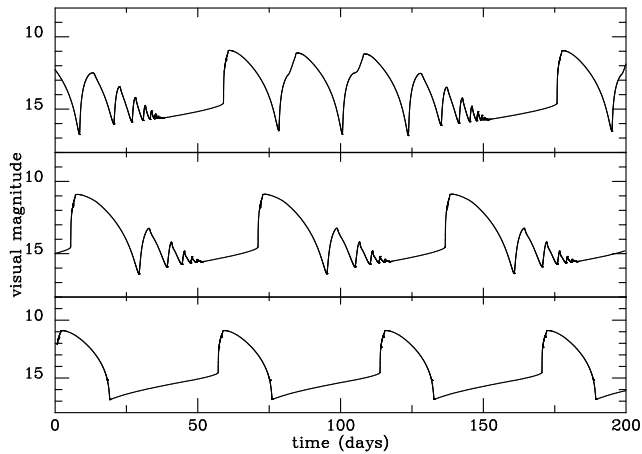,width=\columnwidth}
\caption{Calculated visual magnitude light curves for parameters appropriate
to SS Cyg. In the upper panel, the white dwarf temperature is kept constant,
at a value of 30,000 K; in the intermediate panel, illumination by the
boundary layer is also included (see text), and in the lower panel, we that
the inner parts of the disc are evaporated, and we do not consider the effect
of illumination on the outer disc.}
\label{fig:sscyg}
\end{figure}
\begin{figure}
\psfig{figure=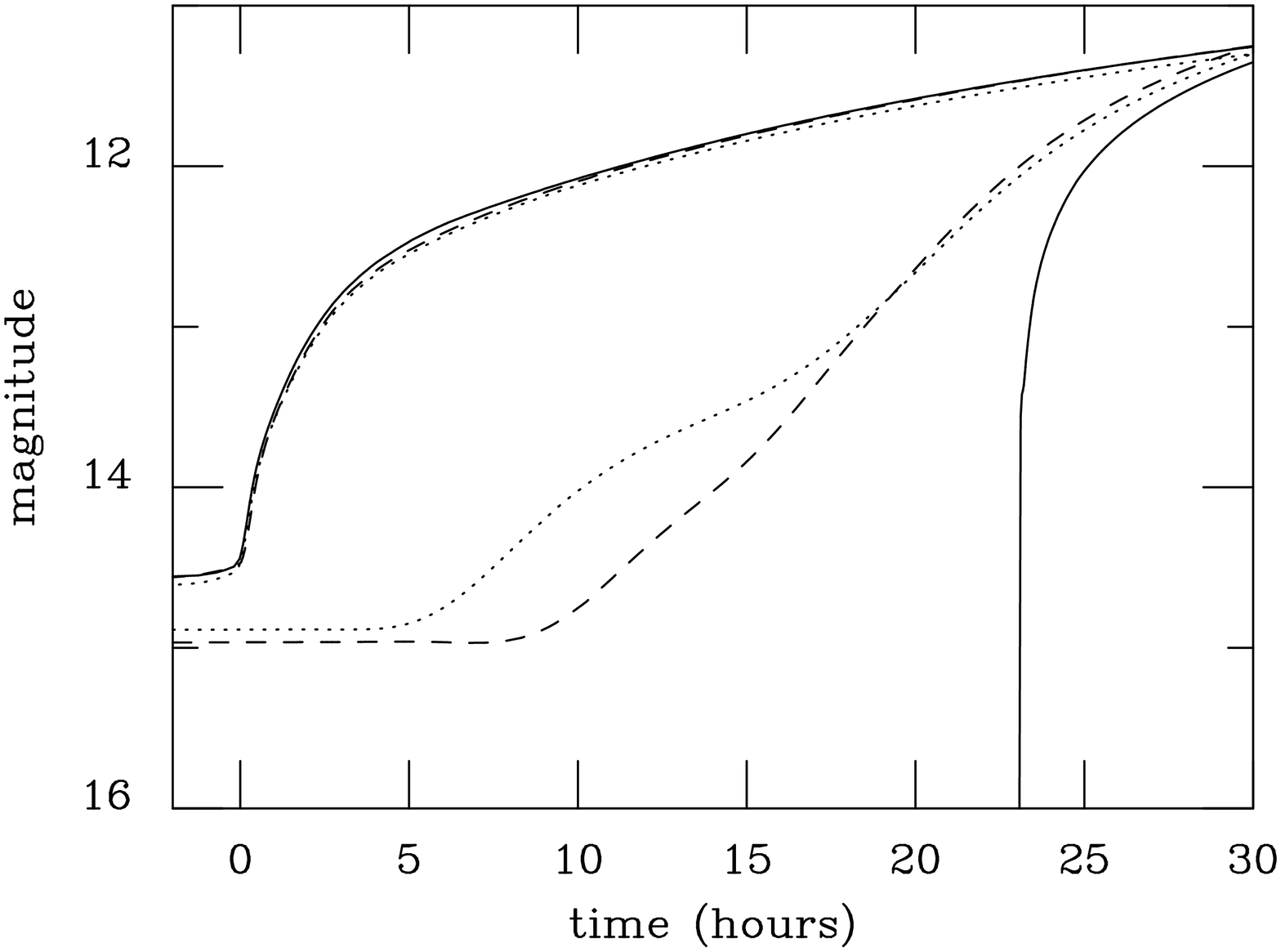,width=\columnwidth}
\caption{Visual (left three curves, almost superimposed) and ``accretion rate"
(right three curves) magnitudes in the case
of SS Cyg. The solid curve is obtained when assuming evaporation of the inner
disc and no illumination; the dashed line corresponds to illumination by both
the white dwarf and the boundary layer, and the dotted line to illumination
by the white dwarf only (see text for parameters)}
\label{fig:sscygrise}
\end{figure}

SS Cyg is a well studied system in which a 1 day EUV delay has been observed
\cite{mau96} that contains a hot, massive white dwarf, whose temperature is
estimated to be in the range 34 -- 40,000 K \cite{w95} for a mass of about
1.2 M$_\odot$ \cite{rk98}. In such a system, the efficiency of accretion is
large, and illumination by the disc itself and by the boundary layer between
the disc and the surface of the white dwarf could be important. The
geometrical extent of the hottest parts of the system is not well known; as a
simplifying assumption, and in order to maximize the effect of accretion
related irradiation, one can assume that the source of illumination is the
white dwarf surface, so that Eq. (\ref{eq:till}) still applies if the
constant temperature $T_*$ is replaced by $T_{\rm s}$ given by:
\begin{equation}
\sigma T_{\rm s}^4 = \sigma T_*^4 + {G M_1 \dot{M} \over 4 \pi R_*^3}
\end{equation}
where $M_1$ is the primary mass. The cooling term $T_{\rm eff}$ has been
calculated for several illumination temperatures, and a cubic spline
interpolation is performed, as described in Dubus et al. \shortcite{dlhc98}.

We show in Fig. \ref{fig:sscyg} the calculated light curves for parameters
appropriate to the case of SS Cyg. We have taken $\alpha_c = 0.02$, $\alpha_h
= 0.10$, $\dot{M} = 1.5 10^{17}$ g s$^{-1}$, $T_*$ = 30,000 K, $\beta$ = 0.5,
$M_1$ = 1.2 M$_\odot$ and $R_* = 5 \times 10^8$ cm. We have also calculated
light curves for $T_*$ = 40,000 K, which are quite similar to those shown
here. A zoom of the outburst rise is shown in Fig. \ref{fig:sscygrise}. For
comparison, we have also shown a case in which illumination by the boundary
layer has been neglected, and a case in which no illumination is assumed, but
matter from the disc is evaporated at a rate: \begin{equation}
\dot{\Sigma}_{\rm ev} = {3 \times 10^{-2} \over (1 + 250 R_{\rm 10}^2)^2}
\;\; \rm g\;s^{-1}cm^{-2} \label{eq:evap} \end{equation} which has a similar
form to the rate used in Hameury et al. \shortcite{hlmn97} (but not the same
numerical coefficients).  $R_{\rm 10}$ is the radius in units of $10^{10}$
cm. In the present case, the radius of the inner edge of the disc is taken to
be about $3.5 \times 10^9$ cm in quiescence, i.e. 7 white dwarf radii; it has
been taken such as to produce a one day delay between the onset of the
outburst in optical and the time at which the disc reaches the white dwarf
surface. One should note that the radius of the inner edge of the disc in
quiescence is very similar to that required to account for the long
recurrence time of WZ Sge if standard values of $\alpha$ are assumed
\cite{hlh97}. It should finally be mentioned that Eq. (\ref{eq:evap}) is
largely empirical, and should be supported by models. Such models exist
\cite{mm94,lmm98}, but the uncertainties are large.

As one can see, illumination by the white dwarf and/or by the boundary
layer also produce light curves which show many small outbursts in
between major ones, and cannot therefore account for the observed light
curves. The situation is slightly better when the boundary layer is
included, but is far from being satisfactory. Similarly, the predicted
EUV delays are too short, since too small a fraction of the disc is
affected by illumination.

By contrast, both the light curves and the EUV delay obtained when one
assumes that the disc is truncated are in good agreement with
observations. The very significant difference between the case of a
hole in the disc and a depletion of the inner regions is precisely the
existence, in the latter case, of a hot, optically thick, central
region that is able to interact with the outer disc and destabilize
it.

One should finally note that the alternation of various types of outbursts in
SS Cyg cannot be explained if no parameter (e.g. $\dot{M}$) is allowed to
vary with time. Cannizzo \shortcite{can93} found that regular sequences of
short and long outbursts were a natural outcome of the disc instability model
for low values of the disc inner radius; one should however keep in mind that
all his outbursts were of the inside-out type (because the outer disc radius
was fixed at a given distance), which is very difficult to reconcile with the
existence of UV and EUV delays. We do also obtain such sequences, even with
alternating inside-out and outside-in outbursts (see e.g. fig.
\ref{fig:sscyg}), but the small outbursts are much fainter than the major
ones, in contradiction with observations. The long term light curve of SS Cyg
shows large variations of the mass transfer rate from the secondary on
timescales of years \cite{cm92}; these variations must be taken into account
when trying to interpret the alternation of outside-in and inside-out
outbursts, and we know that changes in $\dot{M}$ induce changes of the
outburst type (see Figs. \ref{fig:prof1} and \ref{fig:prof10}).

\section{Conclusion}

We have shown that efficient illumination of the accretion disc by a hot
white dwarf affects the dwarf novae outbursts in several ways:
\begin{enumerate}
\item it causes the appearance of several small outbursts in between the major
ones. This conclusion is firm and does not depend on the assumed viscosity,
but merely reflects the unavoidable presence in quiescence of an unstable
transition region between the hot and cold regimes.
\item for moderate white dwarf temperatures and mass transfer rates,
outside-in outbursts are replaced by inside-out outbursts
\item when outside-in outbursts are present, the UV delay is increased. This
happens only for $T_* (1-\beta)^{1/4}$ much above 15,000 K.
\end{enumerate}
The proposition of King \shortcite{k97} that illumination is responsible for
the observable UV delay is therefore difficult to maintain since the UV delay
is increased only for very hot white dwarfs, but then it implies the presence
of small outbursts in between major ones that are not observed.

However, as pointed out by King \shortcite{k97}, such hot white dwarfs are
present in some DNs, and their radiation does illuminate the innermost parts
of the accretion disc. We suggest that the fraction $1-\beta$ of the
illumination flux that affects the vertical structure of the disc might be
smaller than previously estimated, as (i) the opacities are a very
sensitive function of temperature when hydrogen is partly ionized (factors of
the order of 2 in the Rosseland opacity for a temperature difference of only
20\%), and (ii) the grazing angle is quite small; both effects reduce
penetration depth of the UV photons, which could be absorbed in regions which
are not optically thick and would not modify the photospheric boundary
condition but instead lead to the formation of a corona above the cool disc.
The determination of the albedo is, however, much more complex than in the
case of illumination by hard X-rays, where electron scattering contributes
significantly to the opacities.

A much more promising explanation is that the inner parts of the disc that
would be affected by illumination do not exist, precisely because of the
formation of a corona that evaporates the central regions of the disc.
Irradiation would indeed accelerate the evaporation process itself, but the
coupling between evaporation and illumination remains to be understood.

One should finally keep in mind the possibility that the disc instability
model does not apply to dwarf nova outbursts, but then the scenario proposed
by King \shortcite{k97} would not apply anyway.

When this work was almost completed we learned of similar study by
Stehle \& King \shortcite{sk98}.

\subsection*{Acknowledgments}

We are grateful to Kristen Menou for helpful discussions. We are
particularly indebted to Andrew King and Rudi Stehle for providing us with
a preliminary version of their paper \cite{sk98} and for suggesting
that we take into account the self-illumination of the disc. We
acknowledge support from the British-French joint research programme
{\it Alliance}.


\begin{thebibliography}{}

\bibitem[\protect\citename{Cannizzo, }1993]{can93}Cannizzo J.K., 1993, in
	Wheeler J., ed, Accretion Disks in Compact Stellar Systems.
	World Scientific, Singapore, p. 6
\bibitem[\protect\citename{Cannizzo, }1998]{can98}Cannizzo J.K., 1998,
	in Howell S.B., Kuulkers E., Woodward C., eds, ASP Conf. Ser. Vol.
	137, Wild stars in the old West: proceedings of the 13th North American
	Workshop on Cataclysmic Variables and Related Objects. Astron. Soc.
	Pac., San Francisco, p. 308
\bibitem[\protect\citename{Cannizzo \& Mattei, }1992]{cm92}Cannizzo J.K.,
	Mattei J.A., 1992, ApJ, 401, 642
\bibitem[\protect\citename{Dubus et al., }1998]{dlhc98}Dubus G., Lasota
	J.-P., Hameury J.-M., Charles P., 1998, MNRAS, submitted
\bibitem[\protect\citename{Duschl, }1989]{d89}Duschl W., 1989, A\&A, 225, 105
\bibitem[\protect\citename{Friedjung, }1985]{f85}Friedjung M., 1985, A\&A,
	146, 366
\bibitem[\protect\citename{Gammie \& Menou, }1998]{gm98} Gammie C.F., Menou K.,
	1998, ApJ, 492, L75
\bibitem[\protect\citename{Hameury, Lasota \& Hur\'e, }1997a]{hlh97} Hameury
	J.-M., Lasota J.-P., Hur\'e J.-M., 1997a, MNRAS, 287, 937
\bibitem[\protect\citename{Hameury et al., } 1997b]
	{hlmn97} Hameury J.-M., Lasota J.-P., McClintock J.E., Narayan, R.
	1997b, ApJ, 489, 234
\bibitem[\protect\citename{Hameury et al., }1998]
	{hmdl98} Hameury J.-M., Menou K., Dubus G., Lasota J.-P., 1998, MNRAS,
	in press
\bibitem[\protect\citename{Hassall et al., }1983]{hps83} Hassall B.J.M.,
	Pringle J.E., Schwarzenberg-Czerny A., Wade R.A., Wheelan J.A.J.,
	Hill P.W., 1983, MNRAS, 216, 353
\bibitem[\protect\citename{Hubeny, }1991]{h91}Hubeny I., 1991,
	in Bertout C., Collin S., Lasota J.-P., Tran Than Van J., eds,
	Structure and Emission Properties of Accretion Disks.
	\'Editions Fronti\`eres, Gif-sur-Yvette, p. 227
\bibitem[\protect\citename{Idan \& Shaviv, }1996]{ish96}
	Idan I., Shaviv G., 1996, MNRAS, 281, 604
\bibitem[\protect\citename{King, }1997]{k97} King A.R., 1997, MNRAS, 288, L16
\bibitem[\protect\citename{Lasota \& Hameury, }1998]{lh98}
	Lasota J.-P., Hameury J.-M., 1998, in Accretion Processes in
	Astrophysical Systems - Some Like it Hot, Proceedings of the 8th
	Annual October Conference in Maryland. AIP, in press
\bibitem[\protect\citename{Lasota et al., }1995]{lhh95} Lasota J.-P.,
	Hameury J.-M., Hur\'e, J.-M., 1995, A\&A, 302, L29
\bibitem[\protect\citename{Liu, Meyer \& Meyer-Hofmeister, }1992]{lmm98} Liu
	B.F., Meyer F., Meyer-Hofmeister E., 1998, A\&A, submitted
\bibitem[\protect\citename{Livio \& Pringle, }1992]{lp92} Livio M., Pringle
	J.E., 1992, MNRAS, 259, 23p
\bibitem[\protect\citename{Mauche, }1996]{mau96} Mauche C. W., 1996,
	in Bowyer S., Bowyer R.F., eds, IAU Coll. 152, Astrophysics in Extreme
	Ultraviolet. Kluwer, Dordrecht, p. 317
\bibitem[\protect\citename{Meyer, }1990]{m90} Meyer F., 1990, Rev. Mod.
	Astron., 3, 1
\bibitem[\protect\citename{Meyer \& Meyer-Hofmeister, }1994]{mm94} Meyer F.,
	Meyer-Hofmeister E., 1994, A\&A, 288, 175
\bibitem[\protect\citename{Mineshige, }1988]{m88}Mineshige S., 1988, A\&A,
	190, 72
\bibitem[\protect\citename{Patterson et al., }1998]{prkk98} Patterson J.,
	Richman H., Kemp J., Mukai K., 1998, PASP, 110, 403
\bibitem[\protect\citename{Ritter \& Kolb, }1998]{rk98} Ritter H., Kolb U.,
	1998, A\&AS, 129, 83
\bibitem[\protect\citename{Smak, }1989]{s89}Smak J.I., 1989, Acta Astronomica,
	39, 201
\bibitem[\protect\citename{Stehle \& King, }1998]{sk98}Stehle R., King A.R.,
	1998, MNRAS, submitted
\bibitem[\protect\citename{Tuchman, Mineshige \& Wheeler, }1990]
	{tmw90}Tuchman Y., Mineshige S., Wheeler J.C., 1990, ApJ, 359, 164
\bibitem[\protect\citename{van Paradijs, }1996]{vp96}van Paradijs J., 1996,
	ApJ, 464, L139
\bibitem[\protect\citename{Warner, }1995]{w95}Warner B., 1995,
	Cataclysmic Variable Stars, CUP, Cambridge
\end{thebibliography}
\end{document}